\documentclass{article}
\usepackage{spconf,multirow,amsmath,graphicx}
\usepackage{tabu}

\title{Global SNR Estimation of speech signals using Entropy and uncertainty Estimates from Dropout Networks}
\name{Rohith Aralikatti, Dilip Kumar Margam, Tanay Sharma, Abhinav Thanda, Shankar M Venkatesan}
\address{Samsung R\&D Institute India, Bangalore \\ \{r.aralikatti,dilip.margam,tanay.sharma,abhinav.t89,s.venkatesan\}@samsung.com}
\begin{document}
%
\maketitle
\begin{abstract}


This paper demonstrates two novel methods to estimate the global SNR of speech signals. 
In both methods, Deep Neural Network-Hidden Markov Model (DNN-HMM) acoustic model used in speech recognition systems is leveraged for the additional task of SNR estimation. 
In the first method, the entropy of the DNN-HMM output is computed. 
Recent work on bayesian deep learning has shown that a DNN-HMM trained with dropout can be used to estimate model uncertainty by approximating it as a deep Gaussian process. 
In the second method, this approximation is used to obtain model uncertainty estimates. Noise specific regressors are used to predict the SNR from the entropy and model uncertainty. 
The DNN-HMM is trained on GRID corpus and tested on different noise profiles from the DEMAND noise database at SNR levels ranging from -10 dB to 30 dB.

\end{abstract}
\section{Introduction}
\label{sec:intro}

Signal-to-noise ratio (SNR) estimation of a signal is an important step in many speech processing techniques such as robust automatic speech recognition (ASR) (\cite{seltzer2013investigation,lee2016two}), speech enhancement (\cite{ephraim1985speech,nemer1999snr}), noise supression and speech detection. 

The global signal-to-noise ratio (SNR) of a signal $x(t)$ in dB is defined as follows. 
\begin{equation*}
 SNR_{dB} (x) = 10 \, \log_{10} \, \frac{Power(s)}{Power(n)}  
\end{equation*}
The signal $x(t)=s(t)+n(t)$ where $s(t)$ represents the clean signal and $n(t)$ is the noise component.

State-of-the-art ASR has achieved very low error rates with the advent of deep learning. However, performance of ASR systems can still be improved in noisy conditions. Robust ASR techniques such as noise-aware training \cite{seltzer2013investigation} and related methods (\cite{xu2014dynamic},\cite{lee2016two}) require an estimate of the noise present in the speech signal. 

Recently, it has been shown that incorporating visual features (extracted from lip movements during speech) can lead to improved word error rates (WER) during noisy environment (\cite{thanda2017multi},\cite{thanda2016audio}). In \cite{scalart1996speech}, both audio and visual modalities are used for speech enhancement. With the proliferation of voice assistants and front facing cameras in smartphones, using visual features to improve ASR seems feasible. This raises the crucial question - when should the camera be turned on to make use of features from the visual modality? In such scenarios, we can benefit from accurate SNR estimation by turning on the camera in noisy environments.  


In this paper, we present two novel methods to estimate the global SNR (at an utterance level) of a speech signal. 
Both methods require training a DNN based speech classifier on noise free audio using alignments generated from a GMM-HMM model trained for ASR. 
The first method estimates SNR by computing the entropy of the DNN's output. 
The second method uses model uncertainty estimates obtained by using dropout during inference as shown in \cite{gal2016dropout}. 
In section \ref{sec:rwork}, we present related work that has been done. 
Section \ref{sec:entropysnr} describes the entropy based SNR estimator. Section \ref{sec:snruncertainty} describes the dropout based SNR estimator. 
Section \ref{sec:experiments} describes the architecture of the network, the training procedure and the experiments done.
Section \ref{sec:results} presents the results of the paper. The final section \ref{sec:conclusion} has the conclusion.

\section{Related Work}
\label{sec:rwork}

SNR estimation has been an active area of research. In \cite{papadopoulos2016long}, the authors use specific handcrafted features such as signal energy, signal variability, pitch and voicing probability to train noise specific regressors that compute SNR of an input signal.
In \cite{kim2008robust}, the amplitude of clean speech is modelled by a gamma distribution and noise is assumed to be normally distributed. SNR is estimated by observing changes to the parameters of the gamma distribution upon addition of noise.

The NIST-SNR measurement tool uses a sequential GMM to model the speech and non-speech parts of a signal to estimate the SNR. In \cite{morales2011pitch}, a voice activity detector (VAD) is used to classify frames as either voiced, unvoiced or silence and the noise spectrum is estimated from this information. After subtracting the noise spectrum from the input signal to obtain the clean signal, SNR is estimated. In \cite{narayanan2012casa}, computational auditory scene analysis is used to estimate speech dominated and noise dominated portions of the signal in order to obtain SNR.

Estimation of instantaneous SNR is also a subtask in many speech enhancement methods (\cite{scalart1996speech,lee2017deep,cohen2005relaxed,ren2008improved}). In \cite{tchorz2003snr}, a neural network is trained to output the SNR in each frequency channel using amplitude modulation spectrogram (AMS) features which are obtained from the input signal. In \cite{martin1993efficient}, the peaks and valleys of the smoothened short time power estimate of a signal are used to estimate the noise power and instantaneous SNR.
\section{Entropy Based SNR Estimation}
\label{sec:entropysnr}

In this method, a neural network which is trained as a part of ASR system to predict the posterior distribution of HMM states is used. 
The Shannon entropy of the posterior distribution is computed. 
In information theory, Shannon entropy is realisation of the average uncertainity of encoding machine. Similarly in our case the posterior distribution obtained from DNN which is trained as a part of ASR system, acts as an encoding distribution for encoding machine. Whenever the feature vector of clean signal is forwarded through DNN it is expected to give meaningful posterior distribution. But when a feature vector of a noisy signal is forwarded through the neural network, the posteriors are expected to be arbitrary, which in most cases lead to higher entropy value. 
This comes from the assumption that addition of noise to the speech signal results in arbitrary features.

Let $\mathbf{F}_i$ denote the $i^{th}$ input frame of utterance $\mathbf{U}$ and $\mathbf{Y}$ (of dimension $d$) denote the output of DNN. The entropy for given input $\mathbf{F}_j$ is computed as shown in equation \ref{equ:model_entropy}.
\small
\begin{align}
 \mathit{H}(\mathbf{F}_j) &= - \sum_{i=0}^{d} P[\mathbf{Y}_i] \log P[\mathbf{Y}_i] \label{equ:model_entropy} \\
 \mathit{Entropy}(\mathbf{U}) &= \sum_{i=0}^{m} \frac{\mathit{H}(\mathbf{F}_i)}{m} \label{equ:global_entropy} \\
 SNR(\mathbf{U}) &= f_1(Entropy(\mathbf{U})) \label{equ:reg1}
\end{align}
\normalsize

Where $P[.]$ denotes softmax activation, $\mathbf{Y}_i$ is $i^{th}$ dimension of $\mathbf{Y}$. 
The average entropy of all input frames for a given utterance is used as a measure of the entropy for an utterance. 
A polynomial regressor $f_1(.)$ is trained on utterance level entropy values to predict the SNR of speech signal.
The advantage of this method is that it can work on any kind of noise which can randomize the speech signal. 
The DNN-HMM based ASR systems which are sensitive to noisy conditions, can take advantage of entropy values to estimate the SNR with low computational overhead.
In figure \ref{fig:entropy_final}, it is clearly seen that with increase in noise, the average entropy increases. 




\section{SNR Estimation Using Dropout Uncertainty}
\label{sec:snruncertainty}

\subsection{Bayesian uncertainty using dropout}
\label{ssec:bayes}

Gal and Ghahramani showed in \cite{gal2016dropout} that the use of dropout while training DNNs can be thought of as a bayesian approximation of a deep Gaussian process (GP). Using the above GP approximation, estimates for the model uncertainty of DNNs trained using dropout are derived. More specifically, it is shown that uncertainty of the DNN output for a given input can be approximated by computing the variance of multiple output samples obtained by using dropout during inference. The use of dropout during inference, results in different output every time the forward pass is done, for a given input. The variance of these output samples is the uncertainty for the given input.

The above method is used to obtain uncertainty estimates for the DNN that was trained as a part of DNN-HMM based ASR system as explained in section \ref{sec:results}. This DNN is referred as dropout network through out this paper.
If the input is corrupted by noise, it is expected that the model uncertainty derived from dropout will be higher. 
The model uncertainty for given input $\mathbf{F}_j$ is computed as shown in equation \ref{equ:model_uncertainty}.
\small
\begin{align}
 MU(\mathbf{F}) &= \sum_{i=0}^{d} Var[\mathbf{Y}_i] \label{equ:model_uncertainty} \\
 uncertainty(\mathbf{U}) &= \sum_{i=0}^{m} \frac{MU(\mathbf{F}_i)}{m} \label{equ:global_uncertainty} \\
 SNR(\mathbf{U}) &= f_2(uncertainty(\mathbf{U})) \label{equ:reg2} \\
 SNR(\mathbf{U}) &= f_3(uncertainty(\mathbf{U}),Entropy(\mathbf{U})) \label{equ:reg3}
\end{align}
\normalsize
Where $MU$ stands for model uncertainty per frame. The average variance over all input frames is used as a measure of uncertanity for an utterance. 
The SNR of the utterance is estimated as shown in equation \ref{equ:reg2}, where $f_2(.)$ is polynomial regressor trained to predict SNR from uncertainty value.
The the regressor $f_3(.)$ is trained on both uncertainty and entropy of utterance to output SNR value. We have compared the performance of all three regressors in table \ref{tab:result}.

\subsection{Fast dropout uncertainty estimation}
\label{ssec:fdropout}

It may not always be feasible to run the forward pass multiple times per input frame in order to obtain output samples. 
Given the input frame and the weights of the dropout network, it should be possible to algebraically derive the variance and expectation of the output layer.

The uncertainty of the model is the consequence of uncertainty added because of dropout in each layer of network. Following equations depicts how the uncertainity of model can be computed mathematically.
For mathematical simplicity let us consider the neural network with one layer. 
The output of the one layer network with ReLU activation function is: $\mathbf{Y} = ReLU(\mathbf{W} \cdot (\mathbf{D} \circ \mathbf{F})+ \mathbf{b})$. 
Where $\circ$ denotes hadamard product, $\mathbf{D}$ denotes the dropout mask. The variance of $i^{th}$ dimension of output is given as shown in equation \ref{equ:one}.
\small
\begin{align}
Var[\mathbf{Y}_i] &= Var[ReLU(\mathbf{W}_i^{T} (\mathbf{D} \circ \mathbf{F})+ \mathbf{b}))] \nonumber \\
		  &= Var[ReLU(\sum_{j=0}^{m-1} \mathbf{W}_{ij} \mathbf{D}_j \mathbf{F}_j )] \label{equ:one} \\
		  &= Var[ReLU(A_i)] \nonumber
\end{align}
\normalsize
 Where $ A_i = \sum_{j=0}^{m-1} \mathbf{W}_{ij} \mathbf{D}_j \mathbf{F}_j$. $\mathbf{W}_i$ denote $i^{th}$ row of matrix $\mathbf{W}$, $m$ is the dimension of $F$. 
The dropout variable $\mathbf{D}_i$ being a bernoulli variable with probability of success $p$, $Var[D_i] = p(1-p)$. 
\small
\begin{align}
 Var[A_i] &= \sum_{j=0}^{m-1} \mathbf{W}_{ij}^2 \mathbf{F}_j^2 Var[D_j] \nonumber \\
	  &= p(1-p) \sum_{j=0}^{m-1} \mathbf{W}_{ij}^2 \mathbf{F}_j^2 \label{equ:three}
\end{align}
\normalsize
Since all the dropout bernoulli random variables are independent of one another, the equation \ref{equ:three} follows.
 The difficulty comes in computing the $Var[\mathbf{Y}_i]$ because it involves a non-linear ($Relu$) activation function. 
To compute the $Var[\mathbf{Y}_i]$ one has to integrate the $\mathbf{Y}_i$s over all possible dropout distributions ($2^m$ possibilities), which will increase the computational complexity. 
One can proceed from here using the Taylor first order approximation of $m$ variables. 
In \cite{wang2013fast} it is assumed that sum of activation values follows normal distribution following the central limit theorem, but this assumption did not hold good empirically in our case because of multiple layers in network. 

However, the variance of the output is some complex non-linear function of the input and the dropout network weights. 
Therefore it must be possible to train another DNN to learn this non-linear relationship so that the uncertainty can be estimated by a single forward pass of this second network. 
This second neural network from now on will be referred to as the \textit{variance network} in this paper. The variance network explained in section \ref{sssec:variancennet} was able to succesfully learn the mapping from the input frame to the output (dropout uncertainty), as shown in the figure \ref{fig:network2_final}.

%
%

\section{Experiments}
\label{sec:experiments}

A DNN-HMM based ASR system is trained on the Grid corpus \cite{cooke2006audio} ($95\%$ of it is used for training, $5\%$ for testing), which has $34$ speakers and $1000$ utterances per speaker. 
The Mel scale filter-bank features of $40$ dimension, with $5$ contextual frames on both sides are used as input features. 
The duration of $25$ ms and shift of $10$ ms is used in feature extraction process. The activation function used is ReLU, along with dropout with $p=0.2$ ($p$ is probability of dropping a neuron) is used in all hidden layers. 
The output of DNN is of dimension $1415$ corresponding to number of HMM states. 
There are six hidden layers with $1024$ neurons in each layer. 
This DNN which is also reffered to as \textit{dropout network} in this paper is used for estimating entropy and variance in all our experiments, except for section \ref{ssec:fdropout}.


\subsection{Entropy method and dropout uncertainty method}

\begin{figure*}[htb]

\begin{minipage}[b]{0.33\linewidth}
  \centering
  \centerline{\includegraphics[width=6.0cm]{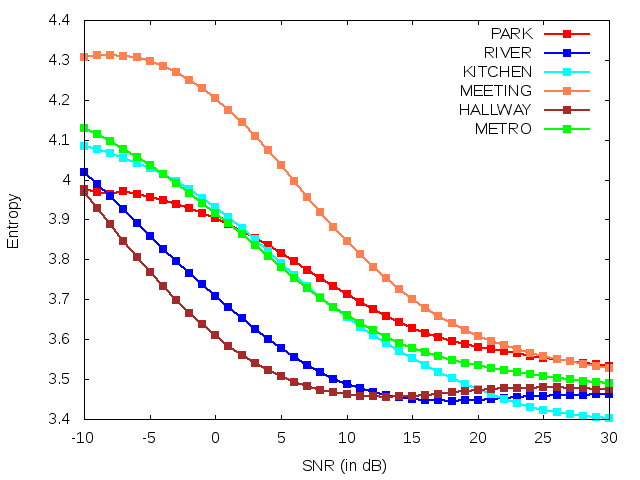}}
  \caption{\small Plot depicts the relationship between averaged entropy of utterance (defined in equation \ref{equ:global_entropy}) with SNR value of utterance for test utterances for six arbitrarily chosen noise types.}\medskip
 \label{fig:entropy_final}
\end{minipage}
\begin{minipage}[b]{0.33\linewidth}
  \centering
  \centerline{\includegraphics[width=6.0cm]{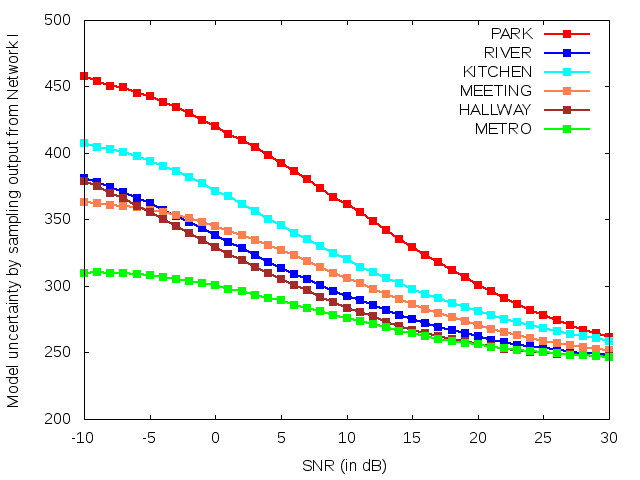}}
  \caption{\small Figure shows the relationship between averaged uncertainity of utterance (as in equation \ref{equ:global_uncertainty}) and SNR value of utterance for test utterances for six arbitrarily chosen noise types.}\medskip
 \label{fig:dropout_final}
\end{minipage}
\hfill
\begin{minipage}[b]{0.33\linewidth}
  \centering
  \centerline{\includegraphics[width=6.0cm]{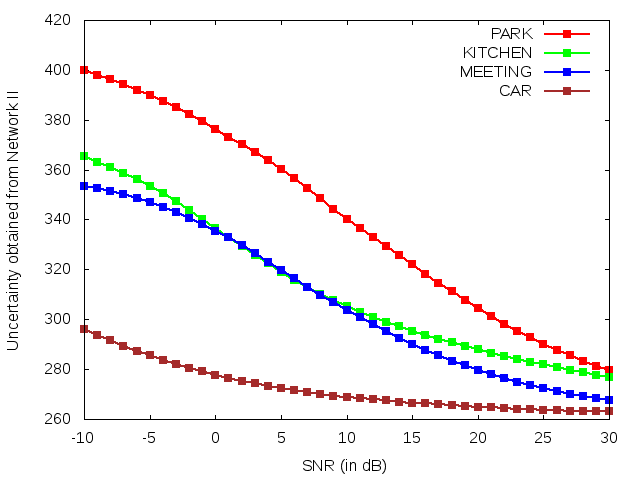}}
\caption{\small Figure shows the relationship between output of variance network and noisy input speech with different SNR values for four unseen (not used in training) noises.}\medskip
 \label{fig:network2_final}
\end{minipage}
%
%
\end{figure*}

We experimented on 16 different noise types from the DEMAND noise dataset, where noise is added to the test set of utterances. We observe that there is a strong correlation between average entropy and SNR as shown in figure \ref{fig:entropy_final}. Similar kind of results for average dropout uncertainty estimates versus the SNR are obtained, where model uncertainty increases with increase in noise as shown in the figure \ref{fig:dropout_final}. 
The variance has been computed by taking 100 output samples per input frame, but we obtained similar results when we reduced the number of samples to 20 per input frame. 
Figure \ref{fig:dropout_final} shows the variation in model uncertainty with respect to SNR for same six arbitrarily chosen noises as in figure \ref{fig:entropy_final}.


The variance computation was done on the output samples obtained from the DNN before the application of softmax to obtain probabilities. This gave better results, since the softmax function tends to exponentially squash the outputs to lie between $0$ and $1$ and this causes the variance along many of the dimensions of the output to be ignored.   

Using the ReLU non-linearity also gave better results as compared to the sigmoid and tanh non-linearity. This is expected, as both the sigmoid and tanh tend to saturate and this does not allow the variance (or model uncertainty) to be propagated to the output layer.

\subsubsection{Variance network (fast dropout uncertainty estimation)}
\label{sssec:variancennet}
This is the network used for fast dropout uncertainty estimation.
The variance network is trained on uncertainty estimates obtained from the dropout network.
 The training is done on utterances from the GRID corpus mixed with noise from the DEMAND \cite{thiemanndemand} dataset using the previously trained dropout network. 
The training is done on utterances mixed with 12 types of noise at 40 different SNR levels (from -10 dB to 30 dB). 
The testing is done on different utterances from the GRID corpus mixed with noise samples not exposed to the network during training.

Variance network is able to succesfully learn the mapping from the input frame to the output uncertainty. The plots shown in the figure \ref{fig:network2_final} shows the variation of output uncertainty for the four types of noise (CAR, PARK, KITCHEN, MEETING) which were not used during training.

%
%
%

\section{Results}
\label{sec:results}

To obtain the SNR of an input signal, we have trained noise specific linear regressors to obtain the SNR value given the uncertainty obtained from variance network and/or entropy. The mean-absolute-error (MAE) for three different types of noise at different SNR levels are shown in Table \ref{tab:result}.

We have compared the result of the three regressors ($f_1$,$f_2$ and $f_3$) described previously with well known SNR estimation methods, namely the NIST STNR estimation tool and the WADA SNR estimation method described in \cite{morales2011pitch}. It is observed that the regressor trained on dropout uncertainty performed better than the entropy based regressor. Indeed, it is observed that the regressor trained on both the dropout uncertainty and entropy perfomed worse than just regressing on the network uncertainty. However, all three regressors have produced better SNR estimates than either WADA or NIST, partiicularly at low SNR levels.

\begin{table}
\centering
\caption{The Mean Absolute Error (MAE) of our SNR estimation methods is compared against pre-existing methods}
\label{tab:result}
\begin{tabular}{lllllll} \hline
Noise & Method & \multicolumn{5}{c}{SNR (dB)} \\ 
type & & -10 & -5 & 0 & 5 & 10 \\ \hline \hline 
\multirow{5}{*}{\rotatebox[origin=c]{90}{DKITCHEN}} & {\small NIST}  & 15.55  & 10.58  & 6.66  & 5.08  & 4.73  \\ 
 & {\small WADA}  & 9.34  & 5.35  & \textbf{1.31}  & \textbf{0.93}  & \textbf{0.67}   \\ 
 & $f_{1}$  & 8.67  & 8.48  & 7.98  & 6.93  & 7.44   \\ 
 & $f_{2}$  & \textbf{2.58}  & \textbf{2.06}  & 2.73 & 3.37 & 2.7 \\ 
 & $f_{3}$  & 3.08  & 2.85  & 3.57  & 4.34  & 4.02  \\ 
\hline 
\multirow{5}{*}{\rotatebox[origin=c]{90}{NPARK}}  & {\small NIST}  & 17.32  & 12.64  & 8.71  & 6.94  & 6.91 \\ 
 & {\small WADA}  & 7.83  & 4.13  & 2.31  & 1.89  & 2.25 \\ 
 & $f_{1}$  & 7.06  & 6.22  & 5.01  & 4.38  & 4.44 \\ 
 & $f_{2}$  & \textbf{2.34}  & \textbf{2.01}  & \textbf{1.86}  & 1.62  & \textbf{1.28} \\ 
 & $f_{3}$  & 2.43  & 2.11  & 1.9  & \textbf{1.61}  & 1.35 \\ 
\hline 
\multirow{5}{*}{\rotatebox[origin=c]{90}{OMEETING}} & {\small NIST}  & 17.25  & 12.97  & 10.46  & 9.26  & 11.3 \\ 
 & {\small WADA}  & 12.11  & 8.44  & 6.61  & 6.08  & 6.39 \\ 
 & $f_{1}$  & 4.51  & 2.54  & 3.17  & 3.69  & 4.28 \\ 
 & $f_{2}$  & \textbf{1.98}  & 1.46  & 1.79  & \textbf{1.98}  & \textbf{1.95} \\ 
 & $f_{3}$  & 2.32  & \textbf{1.24}  & \textbf{1.68}  & 2.12  & 2.12 \\ 
\hline 
\hline 
\end{tabular}
\end{table}

Though we clearly see a correlation between the entropy/dropout uncertainty and the noise in the signal, to finally obtain the SNR value of the signal we have to train a noise specific regressor on top of the entropy/dropout uncertainty values. 
The possibility of directly predicting SNR independent of the background noise is something that needs further research. 
In \cite{papadopoulos2016long}, the authors propose using a DNN to find out which of the noise types most closely resemble the input and use the corresponding regressor to estimate SNR.

However, since dropout network is trained on clean audio, irrespective of the type of noise in the speech signal, the trend of increasing uncertainty with increasing noise did hold even in unseen noise conditions. 
The variance network, which is trained on specific noise types in order to avoid the computational costs of taking samples during inference, clearly maintained this trend even in unseen noise conditions as shown in figure \ref{fig:network2_final}

\section{Conclusion}
\label{sec:conclusion}

In this paper, we have shown that it is possible to extract useful information from the uncertainty (either from entropy or from bayesian estimates) and predict the SNR of a speech signal. Previous research in deep learning based speech processing has not made use of uncertainty information to the best knowledge of the authors. 
Using the above uncertainty information to better design and improve the performance of current ASR and speech enhancement algorithms will be possible future directions of research.
Another possible improvement that can be done is to investigate the possibility of predicting instantaneous SNR instead of global SNR.  
The methods proposed in this paper for SNR estimation do not impose any conditions on the type of noise corrupting the signal. This leaves open the possibility of applying similar noise estimation techniques to non-speech signals. 


\end{document}